\documentclass{aa}  

\usepackage{graphicx}
\usepackage{txfonts}
\usepackage{lipsum}
\usepackage{subcaption}         
\usepackage{lscape}             
\usepackage{placeins}           
                                
\begin{document}

   \title{The evolution of the SFR-M$_\star$ relation at $0.1<z<4$:
   environmental and morphological dependencies
}

%

   \author{Kaimin He\inst{1}
        \and Ke Shi\inst{1}\fnmsep\thanks{Corresponding author: shike.astroph@gmail.com}
        \and Jun Toshikawa\inst{2}
        \and Xianzhong Zheng\inst{3}
        \and Xiaopeng You\inst{1}
        }

   \institute{School of Physical Science and Technology, Southwest University, Chongqing 400715, China
   \and Astronomical Institute, Tohoku University, 6-3 Aramaki, Aoba, Sendai, Miyagi 980-8578, Japan
   \and Tsung-Dao Lee Institute and Key Laboratory for Particle Physics, Astrophysics and Cosmology, Ministry of Education, Shanghai Jiao Tong University, Shanghai, 201210, China}

   \date{Received December 20, 2025}

  \abstract
   {We present a comprehensive study of the relationship between star formation rate (SFR) and stellar mass (M$_\star$) from $z=0.1$ to $z=4$ using a mass-complete sample of approximately 290,000 galaxies from the COSMOS2020 catalog. We find that the SFR-M$_\star$ relation exhibits a pronounced high-mass decline that becomes increasingly evident at lower redshifts. Examining environmental and morphological dependencies, we find strikingly different patterns. For all galaxies, we find galaxies in high-density environments exhibit suppressed star formation rates at $z<1$ especially at high-mass end, while for star-forming galaxies no apparent environmental effect is found at all redshifts. In contrast, galaxy morphology exerts strong influence on the SFR-M$_\star$ relation at $z<2$, in a sense that early-type galaxies exhibit systematically lower star formation rates at fixed mass compared to spirals and irregulars, with this trend persisting even within the star-forming population. These results suggest that internal structural properties (bulge components in particular) continuously regulate star formation efficiency independently of whether galaxies are classified as active or quiescent, whereas external environmental processes primarily serve as rapid quenching mechanisms that increase the fraction of quiescent galaxies at low redshifts. We attribute the observed high-mass decline of the SFR-M$_\star$ relation to COSMOS2020's superior capability for detecting massive star-forming galaxies undergoing ``morphological quenching'' processes. }


   \maketitle
   \nolinenumbers

\section{Introduction}\label{sec:intro}
Understanding how galaxies form stars and how this process evolves across cosmic time represents one of the fundamental challenges in extragalactic astronomy. Over the past two decades, large multi-wavelength surveys have revealed that star-forming galaxies exhibit a tight correlation between their star formation rates (SFRs) and stellar masses (M$_\star$), commonly referred to as the ``star-forming main sequence'' (SFMS) \citep[e.g.,][]{Daddi07,Elbaz07,Noeske07,Koyama13,Speagle14,Popesso23}. 

The SFMS provides a powerful framework for studying galaxy evolution. Galaxies on the main sequence represent systems in quasi-equilibrium states where gas accretion, star formation, and feedback processes balance to maintain relatively constant specific star formation rates (sSFR$\equiv$SFR/M$_\star$) over extended timescales. Deviations from this relation -- either those above or below the SFMS -- signal galaxies undergoing intense star formation or quenching, driven by mergers, AGN activity, or environmental processes.  The evolution of the main sequence normalization, scatter, and shape across cosmic time therefore encodes crucial information about the physical mechanisms regulating star formation throughout the universe's history. Observations suggest that the normalization increases systematically toward higher redshifts, with star formation rates at fixed stellar mass rising by nearly two orders of magnitude from $z=0$ to $z=4$ \citep{Speagle14}, reflecting increasing molecular gas fraction towards the early Universe \citep{Genzel15,Scoville16}. Furthermore, the intrinsic scatter ($\sim0.3$ dex) reveals the stochasticity in gas accretion histories and variations in star formation efficiency among systems of similar stellar mass \citep{Tacchella16,Caplar19}.

A particularly intriguing feature that has emerged from recent studies is the turnover or flattening of the main sequence at high stellar masses. Unlike simple power-law relations that predict monotonically increasing SFRs with stellar mass, observations reveal that the most massive galaxies (M$_\star$ $\gtrsim10^{10}$ M$_\odot$) exhibit systematically lower specific SFR and reach a asymptotic value. This high-mass turnover has been documented across multiple surveys and redshift ranges, though its detailed shape, evolution with cosmic time, and physical origin remain subjects of active debate \citep[e.g.,][]{Whitaker14,Schreiber15,Lee15,Erfanianfar16,Leslie20,Leja22,Popesso23}. 

The physical mechanisms responsible for this high-mass turnover are likely dominated by internal processes. Massive galaxies, which are predominantly centrals \citep[e.g.,][]{Peng12, Popesso19}, may experience reduced star-forming efficiency due to the growth of the quiescent bulge component \citep{Martig09,Abramson14}, or AGN feedback that heats or expels gas reservoirs \citep[e.g.][and references therein]{Fabian12}. Additionally, in massive dark matter halos ($\gtrsim10^{12} M_\odot$), virial shock heating prevents infalling gas from cooling efficiently onto the galaxy, further suppressing star formation \citep{Dekel06,Daddi22}. While environmental processes such as ram-pressure stripping \citep[e.g.,][]{Gunn72} and strangulation \citep[e.g.,][]{Larson80} can also contribute, these mechanisms are thought to operate primarily on satellite galaxies, which constitute a smaller fraction of the galaxy population at these high masses \citep[e.g.,][]{Popesso19}.

Previous studies of the main sequence have employed various approaches and datasets, each with different advantages and limitations. Many deep surveys optimized for high-redshift studies have relatively small field of view (e.g., GOODS, CANDELS with area $\sim0.1-0.2$ deg$^2$; \cite{Elbaz11, Salmon15,Tomczak16}). While reaching faint magnitudes, these surveys suffer from strong cosmic variance and may miss many rare objects such as massive star-forming galaxies at low redshifts to robustly constrain the high-mass end of the main sequence. On the other hand, wide field surveys such as SDSS \citep{Leslie16,Popesso19} provides excellent samples in the local universe but lack high-redshift samples and multiwavelength coverage to investigate the redshift evolution of the main sequence. In addition, several studies have compiled results from multiple surveys to investigate the main sequence evolution \citep[e.g.,][]{Speagle14,Popesso23}. However, although these studies carefully corrected for systematic differences in various datasets, the intrinsic scatter introduced by different IMFs, SPS models, dust laws, SFR indicator and photometric methods can still have non-negligible effects on the main sequence, making it difficult to distinguish genuine physical trends from systematics in the main sequence.

The Cosmic Evolution Survey (COSMOS) is currently one of the deepest and richest surveys that serves as a premier laboratory for studying galaxy evolution across cosmic time \citep{Scoville07}. Covering approximately 2 square degrees with extensive multiwavelength data from near-UV to mid-infrared, the latest COSMOS2020 catalog \citep{Weaver22} represents a substantial advancement over previous releases. Utilizing state-of-the-art SED fitting techniques, COSMOS2020 provides accurate photometric redshifts and physical properties for nearly one million sources extending to $z\sim10$, enabling unprecedented statistical studies of the SFMS across cosmic history.

In this paper, we present a comprehensive analysis of SFR-M$_\star$ relation in the range  $z\sim0-4$ using COSMOS2020. Our analysis examines both the full galaxy sample and star-forming sample (SFMS in this case), allowing us to assess how the inclusion of quiescent systems affects the derived relation. We also investigate in detail how environment and morphology affect the SFR-M$_\star$ relation. This paper is organized as follows. In Section~\ref{sec2} we describe the COSMOS2020 catalog and our sample selection. In Section~\ref{sec3} we describe our analysis of the main sequence relation, including the functional form employed and the redshift evolution of the fitted parameters. Section~\ref{SFRD} investigates the cosmic star formation rate density derived from our main sequence measurements.  Sections~\ref{env} and \ref{morp} examine environmental and morphological dependencies of the main sequence, respectively. Section~\ref{compare} compares our results with literature measurements. We summarize our results in Section~\ref{sum}. Throughout this paper we use the Planck cosmology from \cite{Planck16}.  We assume a \cite{Chabrier03} initial mass function unless otherwise stated.
\section{Data} \label{sec2}
In this section, we describe the data used in this work as well as our sample selection.

The study of SFR-M$_\star$ relation in this work is based on the COSMOS2020 photometric catalog \citep{Weaver22}. Comparing with the previous version \citep{Laigle16}, the new catalog includes ultra-deep optical data from Hyper Suprime-Cam (HSC) Subaru Strategic Program (SSP) PDR2 \citep[SSP;][]{Aihara19}, the deeper $u^*$ and new $u$ band imaging from the Canada France-Hawaii Telescope (CFHT) program CLAUDS \citep{Sawicki19}, the fourth UltraVISTA data release \citep{Moneti23} reaching one magnitude deeper in $K_S$ band over the entire area, and the inclusion of all \textit{Spitzer} IRAC data in the COSMOS \citep{Moneti22}.

COSMOS2020 includes roughly 966,000 sources, providing accurate measurements of photometric redshift and physical properties of galaxies. The catalogs are generated using two independent photometric methods for source extraction. \textsc{The CLASSIC} catalog performs standard aperture photometry using \textsc{SExtractor} \citep{Bertin96} on PSF-homogenized images for optical/near-infrared bands and uses \textsc{IRACLEAN} software \citep{Hsieh12} to perform photometry on Spitzer/IRAC images. In contrast, \textsc{The FARMER} catalog is produced using a new profile fitting method based on the source modeling code \textsc{The Tractor} \citep{Lang16}. Although the two methods perform equally well and are highly consistent with each other, \textsc{The FARMER} can detect fainter and higher density of high-z sources, likely due to its ability to deblend sources at fainter magnitudes \citep{Weaver22}. In this work, we utilize \textsc{The FARMER} catalog since it provides more high-$z$ sources ($\sim$11\% more at $3<z<4$), benefiting our study of the SFR-M$_\star$ relation at high redshifts.

To create a deep and clean sample, we apply several filtering criteria. First, we use flags that mask out bright stars and edges of the HSC and Suprime-Cam images, and require sources to be in the UltraVISTA regions (FLAG\_COMBINED=0 in \textsc{The FARMER} catalog). Second, to separate galaxies from stars and AGNs, we apply the galaxy classification from the catalog (lp\_type=0), which is derived from the SED-fitting code \textsc{LePhare} \citep{Arnouts99, Ilbert06} using combined morphological and SED criteria. These filtering steps result in a final sample of 711,290 galaxies.

We limit our study to galaxies at $z_\textrm{phot}<4$. There are two primary reasons for this. First, the larger redshift uncertainties at higher redshifts create difficulties in interpreting results. Since the COSMOS2020 catalog derives photometric redshifts using spectral energy distribution (SED) fitting with only optical-MIR data, the lack of FIR-radio information makes it challenging to accurately capture the full spectral range of sources. Consequently, many low-redshift galaxies may be misidentified as high-redshift sources due to confusion between the Balmer break at lower redshift and the Lyman break at higher redshift, an issue already noted in previous studies \citep[e.g.,][]{Dunlop07,Shi191,Zavala23}. Second, in this work we would also like to study the environmental impact on the SFR-M$_\star$ relation. However, at higher redshift the number of galaxies becomes too small which makes it difficult to properly define the environment. For example, less than 10,000 galaxies are found at $4<z<5$ and less than 5,000 galaxies at $z>5$, despite the large field coverage of the COSMOS survey. This is due to the relatively shallow depth of the COSMOS field compared to deeper and smaller surveys such as CANDELS \citep{Grogin11, Koekemoer11}. However, the wider area of COSMOS is essential for robust environmental measurements.

The next step is to select a mass-complete sample to constrain the SFR-M$_\star$ relation more accurately. The COSMOS2020 catalog classifies galaxies into two types: star-forming and quiescent. Following \cite{Ilbert13}, quiescent galaxies are defined to have rest-frame colors $M_\textrm{NUV}-M_r>3(M_r-M_J)+1$ and $M_\textrm{NUV}-M_r>3.1$ in the NUV$-r$ vs. $r-J$ diagram. To ensure our sample is mass-complete, we use the stellar mass completeness limits calculated in \citet{Weaver22}. Their analysis followed procedures introduced by \cite{Pozzetti10}, where masses reported by \textsc{LePhare} are rescaled to match the IRAC channel 1 sensitivity limit $m_\textrm{lim}=26$, and the limiting mass is then taken to be the 95th percentile of the rescaled mass distribution. The mass completeness is determined separately for both star-forming and quiescent galaxies (shown in Figure 20 of \cite{Weaver22}) as a function of redshift. We adopt their mass completeness limits for both galaxy types, removing galaxies with \textsc{LePhare} best-fit mass lower than the completeness limit, resulting in a mass-complete sample of 289,624 galaxies.

We divide the sample into seven redshift bins, with each bin containing at least $\sim$25,000 galaxies to ensure our results are not biased by small number statistics. The sample size of each bin is summarized in Table \ref{table1}. The physical properties, such as stellar masses and SFRs, are derived via SED-fitting techniques (optical-MIR) using \textsc{LePhare} in the COSMOS2020 catalog. We refer readers to \cite{Weaver22} for a detailed description of the methods.

\begin{table}[htbp] 
\centering
\caption{Summary of the mass-complete sample}
\resizebox{\columnwidth}{!}{%
\begin{tabular}{ccc}
\hline
\hline
Redshift Range & Number of Galaxies & Number of Quiescent Galaxies\\
\hline
$0.1\leq z<0.7$ & 44783 & 3455 \\
$0.7\leq z<1.0$ & 47279 & 2838 \\
$1.0\leq z<1.5$ & 68209 & 2672 \\
$1.5\leq z<2.0$ & 46191 & 1327 \\
$2.0\leq z<2.5$ & 30321 & 384 \\
$2.5\leq z<3.0$ & 27271 & 158 \\
$3.0\leq z<4.0$ & 25570 & 323 \\
\hline
\label{table1}
\end{tabular}%
}
\end{table}

\section{The SFR-M$_\star$ relation} \label{sec3}

In this section, we present our measurement of the SFR-M$_\star$ relation in different redshift bins defined in the previous section. We also establish a functional form to further investigate its evolution across cosmic time.

\subsection{Parameterizing the SFR-M$_\star$ relation} \label{param}
The simplest way to quantify the SFR-M$_\star$ 
relation is to use a power-law of the form  ``log~SFR~$=$~$\alpha$~log~M$_\star$~$+$~$\beta$'' where $\alpha$ and $\beta$ are time-dependent constants \citep[e.g.,][]{Speagle14}. However, many studies have challenged this simple power-law and suggested a turnover or flattening at the high-mass end.

\begin{figure*}[h] 
\includegraphics[width=1\textwidth]{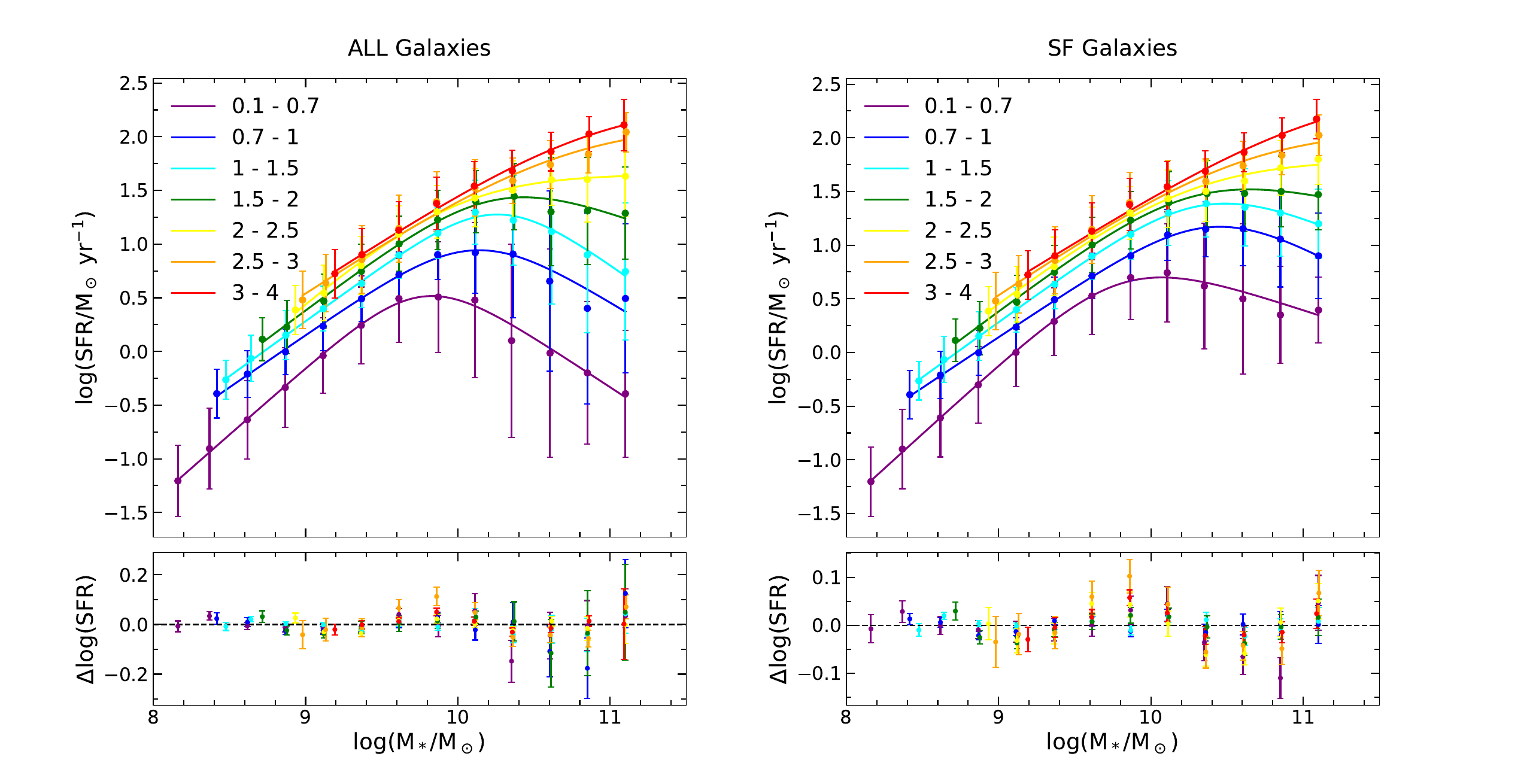}
\caption{
  SFR-M$_\star$ relations for all galaxies (left panel) and star-forming galaxies (right panel). The median SFR in each mass bin are plotted with errorbars denoting the median absolute deviations. Solid lines represent the best-fit relations of Equation~\ref{eq1}. Bottom panels show the residuals calculated as the difference between the measured and best-fit log(SFR) values.
}
\label{figure1}
\end{figure*}

To directly examine the behavior of the SFR-M$_\star$ relation, in Figure~\ref{figure1} we show the median SFR in each mass bin across different redshift ranges. For both all galaxies and star-forming galaxies, the data reveal a clear and systematic departure from simple power-law behavior: a pronounced turnover and decline emerge at high masses for $z<2$. While at higher redshifts ($z>2$), the turnover becomes less pronounced, without apparent decline across the mass range probed by our data.

To better illustrate the non-linear trend of the SFR-M$_\star$ relation, several functional forms have been proposed, such as a broken power law \citep{Whitaker14, Leja22}, a quadratic function \citep{Popesso23}, or a physically motivated functional form introduced by \cite{Lee15} that has been adopted by many subsequent studies \citep[e.g.,][]{Tomczak16,Lee18,Leslie20,Daddi22,Popesso23,Cooke23,Koprowski24}.

However, none of these parameterizations can adequately describe the decline at the high-mass end seen in Figure~\ref{figure1}. In this work we would like to investigate the SFR-M$_\star$ relation using a physically motivated approach. To this end, we modify the functional form of \cite{Lee15} as follows:
\begin{equation} \label{eq1}
\mathrm{log(SFR)=S_0-log[(10^{M_\star}/10^{M_0})^{-\gamma_1}+(10^{M_\star}/10^{M_0})^{\gamma_2}]},
\end{equation}
where M$_\star$ and M$_0$ are in log scale.
This function is a bell-shaped curve that rises to its peak around M$_0$ and then declines to zero as M$_\star$ approaches infinity. S$_0$ is the scaling factor which is proportional to the peak of the function, whereas M$_0$ determines approximately the location at which the turnover occurs. $\gamma_1$ and $\gamma_2$ (both positive) denote the slopes at low and high masses, respectively. We note that this function reduces to the \cite{Lee15} parameterization when $\gamma_2=0$. 

As seen in Figure~\ref{figure1}, the SFR-M$_\star$ relation exhibits different behaviors at $z<2$ and $z>2$. At $z<2$, the SFR declines after it reaches its peak; whereas at $z>2$, it asymptotically approaches to a maximum value. Also at $z>2$, we note that the parameter uncertainties become very large if we allow all the parameters to vary. Thus to obtain better constraints, we fit the relation using Equation~\ref{eq1} with different parameter settings for these two redshift regimes. For $z<2$, all four parameters -- S$_0$, M$_0$, $\gamma_1$, and $\gamma_2$ -- are allowed to vary. For $z>2$, we set $\gamma_2=0$ so that the function reduces to the \cite{Lee15} form. 

We fit Equation~\ref{eq1} to the median values of SFR and M$_\star$ in each redshift bin using the Python package \textsc{CURVE\_FIT}.  We perform this fitting for both the full galaxy sample and the star-forming galaxy subsample. The fitting results are presented in Table~\ref{table2}. We further validate our model choice by comparing it with the \cite{Lee15} model using the Bayesian Information Criterion \citep[BIC;][]{Schwarz78}, a widely adopted tool for model selection in astrophysics and cosmology \citep[e.g.,][]{Liddle07, Trotta08, Shi12}. Assuming a Gaussian posterior distribution, the BIC can be expressed as $\mathrm{BIC} = \chi^2 + k \ln N$, where $k$ is the number of free parameters and $N$ is the number of data points. When comparing different models, ones with lower BIC values are preferred. At $z<2$, our model yields significantly lower BIC values than the \cite{Lee15} model, with $\Delta$BIC$\gg$10, providing very strong statistical evidence in favor of our parameterization \citep{Burnham02}. 

Figure~\ref{figure1} shows the best-fit curves of the SFR-M$_\star$ relation for all galaxies and star-forming galaxies across our redshift range. The median SFR in each mass bin are plotted for each redshift bin, with errorbars denoting the median absolute deviations of the SFR within each bin. Our parameterization fits the data fairly well, as evidenced by the consistently small residuals shown in the bottom panels, which typically remain within $\pm$0.2 dex of the fitted relation. The overall trends for all galaxies and star-forming galaxies are similar, which is not surprising since star-forming galaxies dominate the full sample (Table~\ref{table1}). However, a notable difference emerges at higher masses, where the full sample exhibits a steeper decline compared to the star-forming subsample. This difference suggests that the inclusion of quiescent galaxies, which have systematically lower SFRs, enhances the observed turnover at the high-mass end of the relation.

Figure~\ref{figure1} also clearly shows that the turnover is not simply due to the presence of quiescent galaxies, that even actively star-forming galaxies experience reduced star formation efficiency at the highest stellar masses. We will discuss the possible physical mechanism responsible for this high-mass decline in detail in Section~\ref{disc}.

\begin{table*}[h] 
\centering
\caption{Best-fit parameters of the SFR-M$_\star$ relation}
\begin{tabular}{c|cccc|cccc}
\hline
\hline
\multicolumn{5}{c|}{All Galaxies} & \multicolumn{4}{|c}{Star-forming Galaxies} \\
\hline
Redshift Range & M$_0$ & S$_0$ & $\gamma_1$ & $\gamma_2$ & M$_0$ & S$_0$ & $\gamma_1$ & $\gamma_2$\\
\hline
$0.1\leq z<0.7$ & 9.77$\pm$0.04 & 0.81$\pm$0.03 & 1.25$\pm$0.03 & 0.93$\pm$0.06 & 9.82$\pm$0.06 & 0.95$\pm$0.04 & 1.29$\pm$0.04 & 0.47$\pm$0.07\\
$0.7\leq z<1.0$ & 10.11$\pm$0.07 & 1.24$\pm$0.04 & 0.98$\pm$0.04 & 0.9$\pm$0.1 & 10.41$\pm$0.03 & 1.47$\pm$0.01 & 0.94$\pm$0.01 & 0.79$\pm$0.07\\
$1.0\leq z<1.5$ & 10.26$\pm$0.04 & 1.58$\pm$0.02 & 1.03$\pm$0.02 & 1.0$\pm$0.1 & 10.35$\pm$0.03 & 1.67$\pm$0.02 & 1.03$\pm$0.01 & 0.61$\pm$0.06\\
$1.5\leq z<2.0$ & 10.3$\pm$0.1 & 1.71$\pm$0.05 & 1.06$\pm$0.05 & 0.5$\pm$0.2 & 10.3$\pm$0.1 & 1.76$\pm$0.05 & 1.06$\pm$0.04 & 0.3$\pm$0.1\\
$2.0\leq z<2.5$ & 9.97$\pm$0.05 & 1.65$\pm$0.02 & 1.22$\pm$0.05 & 0 & 10.2$\pm$0.1 & 1.80$\pm$0.04 & 1.08$\pm$0.07 & 0\\
$2.5\leq z<3.0$ & 10.7$\pm$0.2 & 2.1$\pm$0.1 & 0.93$\pm$0.09 & 0 & 10.6$\pm$0.2 & 2.1$\pm$0.1 & 0.96$\pm$0.09 & 0\\
$3.0\leq z<4.0$ & 10.9$\pm$0.1 & 2.33$\pm$0.07 & 0.93$\pm$0.05 & 0 & 11.1$\pm$0.2 & 2.5$\pm$0.1 & 0.89$\pm$0.05 & 0\\
\hline
\multicolumn{9}{p{17cm}}{Notes. Best-fit parameters of Equation~\ref{eq1} at different redshifts, for all galaxies and star-forming galaxies respectively. $\gamma_2$ is fixed to zero at $z>2$.} \\
\end{tabular}
\label{table2}
\end{table*}

\subsection{Redshift evolution of model parameters}
\begin{figure}[h] 
    \centering
    \includegraphics[width=0.5\textwidth]{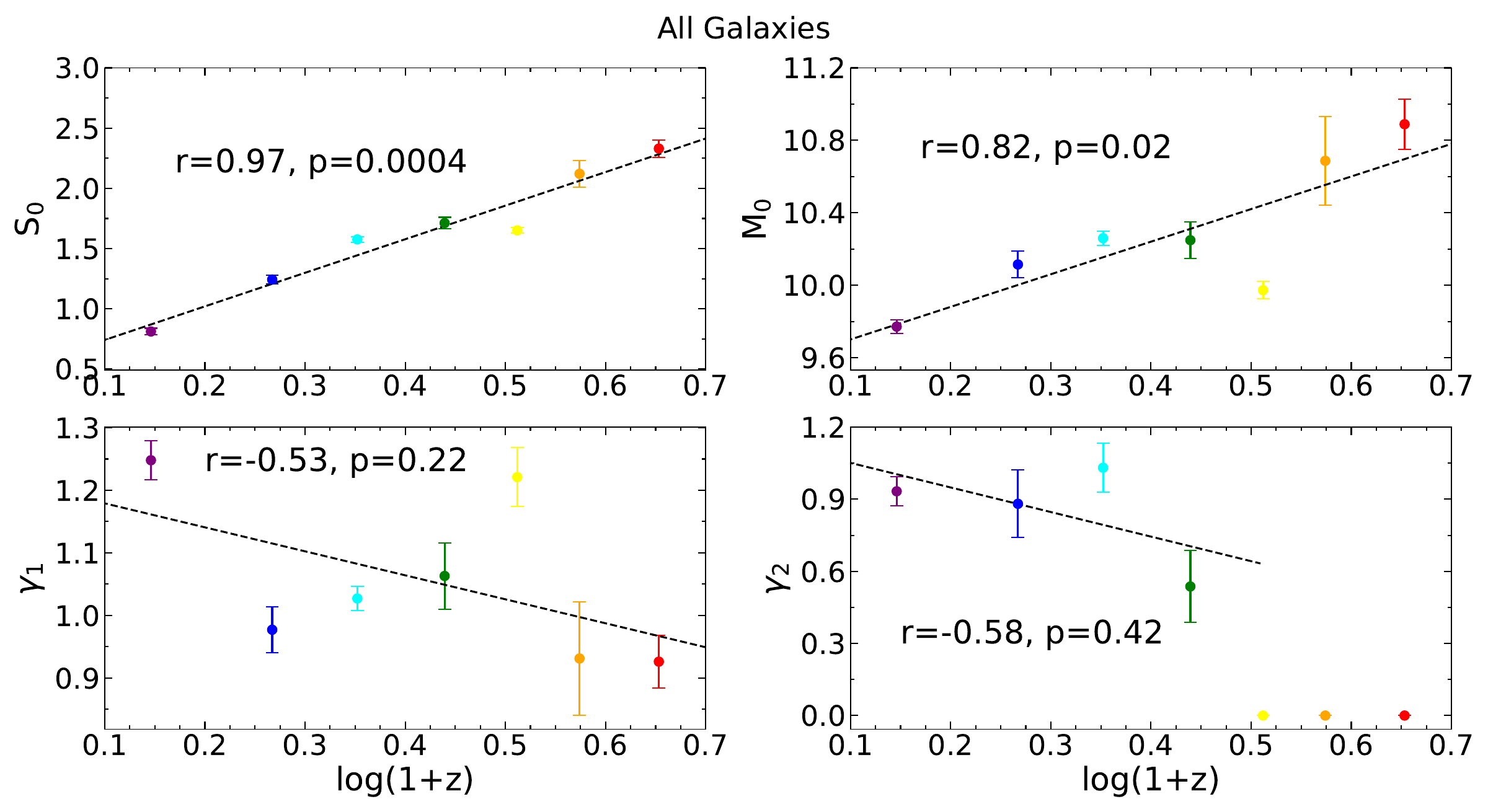}
    \includegraphics[width=0.5\textwidth]{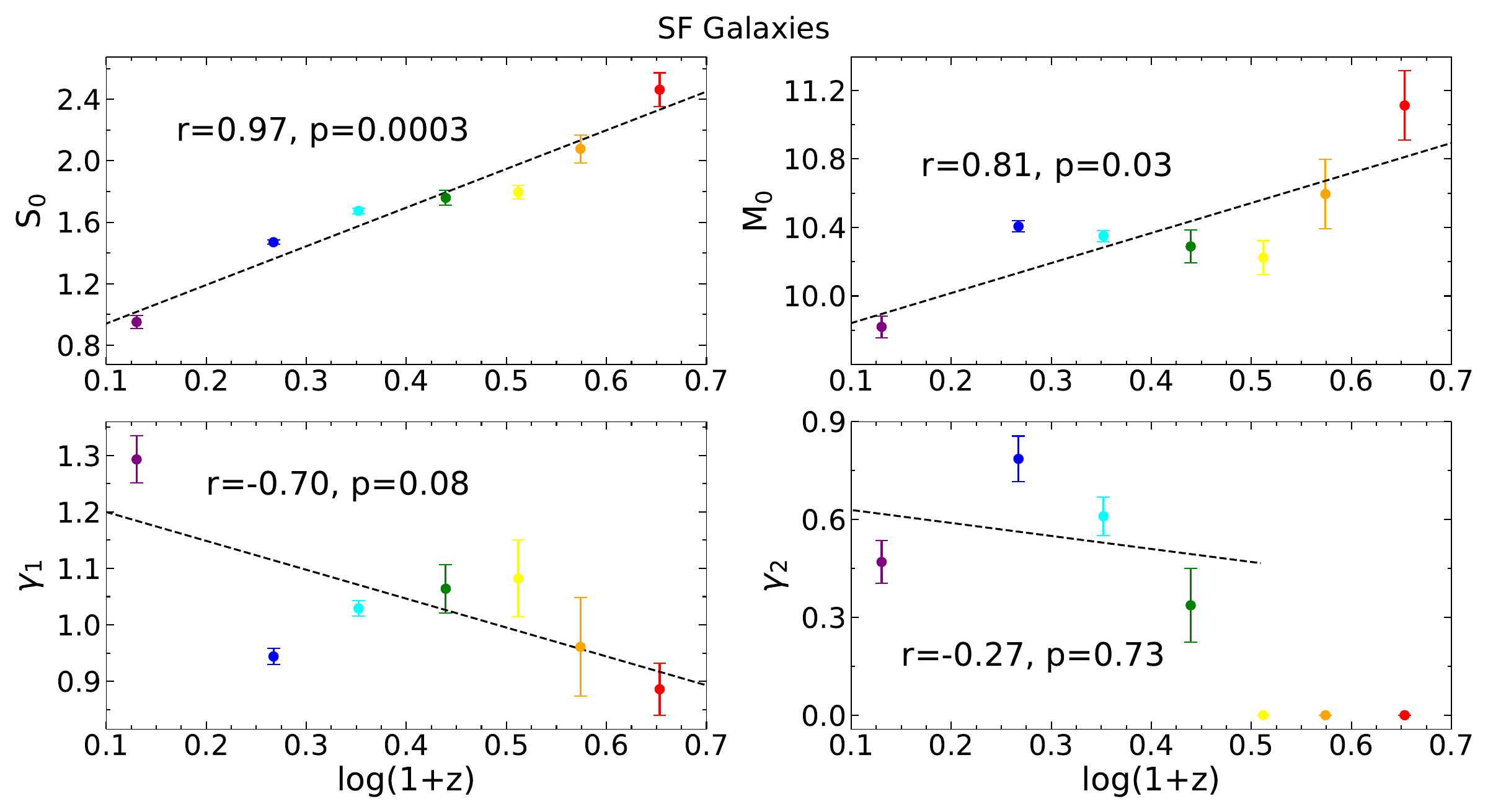}
    \caption{Redshift evolution of the best-fit parameters for all galaxies (upper panel) and star-forming galaxies (lower panel). For each parameter, we perform a linear fit shown as the dotted line. Note that $\gamma_2$ is fixed to zero at $z>2$ and not included in the fit. The inset text shows the Pearson correlation coefficient $r$ and the corresponding $p$-value.}
\label{figure2}
\end{figure}

Table~\ref{table2} shows the best-fit parameters of Equation~\ref{eq1} in different redshift bins. To investigate the time dependence of these parameters, we plot the best-fit values as a function of redshift in Figure~\ref{figure2}. This analysis allows us to quantify how the shape, normalization, and characteristic turnover mass of the main sequence change across different cosmic epochs.

As shown in Figure~\ref{figure2}, there are strong linear correlations between both S$_0$ and log($1+z$), and between M$_0$ and log($1+z$). In contrast, the trends for $\gamma_1$ and $\gamma_2$ are less clear, showing considerable scatter and no obvious systematic evolution with redshift. To quantify these relationships statistically, we calculate the Pearson correlation coefficient for each parameter. The correlation coefficients and the corresponding $p$-values are shown in the figure for each parameter. The small $p$-values of S$_0$ and M$_0$ ($<$0.05) confirm statistically significant linear relationships with cosmic time. However, for $\gamma_1$ and $\gamma_2$, the $p$-values are much larger than 0.05, indicating very weak or negligible correlations with redshift. A similar trend has been reported by \cite{Lee15} and \cite{Popesso23}, who also found $\gamma_1$ to be nearly independent of redshift.  This may reflect the fact that the processes governing star formation efficiency -- such as gas cooling, stellar feedback, and AGN feedback -- scale primarily with stellar mass rather than cosmic epoch. 

We also perform linear fits to the four parameters, shown as the dotted lines in Figure~\ref{figure2}. For the parameters that exhibit significant correlations with redshift (i.e., S$_0$ and M$_0$), we obtain the following relationships: $S_0=(2.95\pm0.35)\times\textrm{log}(1+z)+(0.43\pm0.16)$ for all galaxies, and $S_0=(2.52\pm0.28)\times\textrm{log}(1+z)+(0.69\pm0.13)$ for star-forming galaxies; while $M_0=(1.76\pm0.55)\times\textrm{log}(1+z)+(9.54\pm0.25)$ for all galaxies and $M_0=(1.75\pm0.57)\times\textrm{log}(1+z)+(9.67\pm0.26)$ for star-forming galaxies. 

The similar slopes of S$_0$ in both samples (2.95 vs. 2.52) indicate that the main sequence normalization evolves at comparable rates regardless of whether quiescent galaxies are included, while the positive slopes confirm that star formation rates were systematically higher at earlier cosmic times. For M$_0$, the nearly identical slopes (1.76 vs. 1.75) suggest that the characteristic turnover mass evolves consistently across both galaxy populations, with higher turnover masses occurring at higher redshifts. Our results are in good agreement with \cite{Lee15} and \cite{Popesso23}, who also found tight linear correlations for the main sequence normalization and the turnover mass with cosmic time. On the other hand, $\gamma_1$ and $\gamma_2$ show no obvious time evolution. Yet we note that although the low-mass slope $\gamma_1$ is similar for both samples at different redshifts, the high-mass slope $\gamma_2$ of all galaxies is apparently larger than that of star-forming galaxies at $z<2$, indicating a more rapid decline of SFR for all galaxies at high-mass end, primarily driven by quiescent galaxies within.

\section{Discussion} \label{disc}
\subsection{The evolution of cosmic SFR density} \label{SFRD} 
The cosmic star-formation rate density (SFRD) represents one of the most fundamental observables in extragalactic astronomy, quantifying the rate at which the Universe as a whole is forming stars at any given epoch. Observations show that the cosmic SFRD rose steadily to its peak at cosmic noon ($z\sim2-3$), then declined by nearly an order of magnitude to the present day \citep[i.e.,][]{Madau14}, reflecting fundamental changes in the physical processes governing star formation and galaxy evolution. In this section, we investigate the cosmic SFRD using the state-of-the-art COSMOS2020 catalog.

\begin{figure*}[htbp]
	\centering
    \includegraphics[width=1\textwidth]{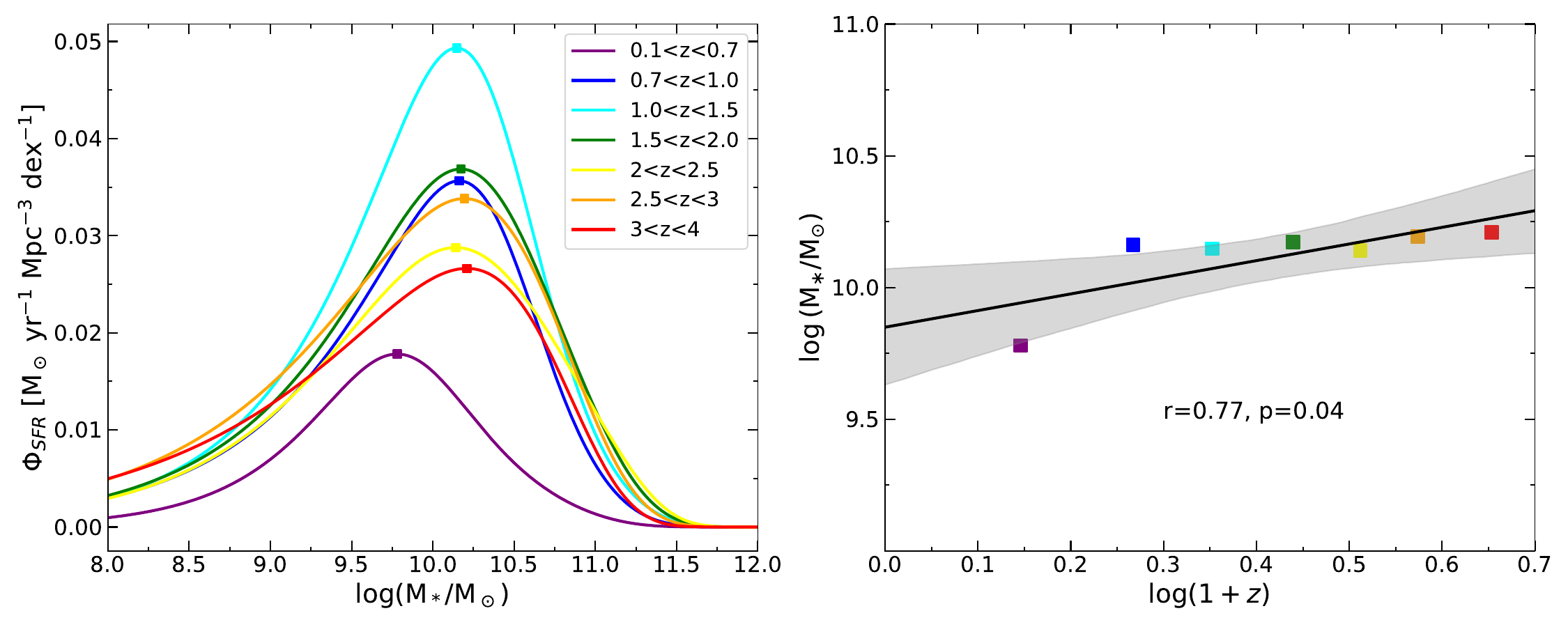}
    \caption{Left: the cosmic SFRD as a function of stellar mass at different redshifts. The square point indicates the characteristic stellar mass at which the SFRD reaches to its peak. Right:  time evolution of the characteristic stellar mass, where the solid line is our linear fit while the shaded area denotes the 1$\sigma$ confidence level. The corresponding Pearson coefficient $r$ and $p$-value are also shown.}
    \label{figure3}
\end{figure*}

We measure the cosmic SFRD by integrating the galaxy stellar mass function from the COSMOS2020 catalog \citep{Weaver23}. First, we calculate the SFRD as a function of stellar mass by multiplying the best-fit SFR from Eq.~\ref{eq1} and Table~\ref{table2} with the stellar mass function of star-forming galaxies from \cite{Weaver23} in each redshift bin.  Our results are shown in the left panel of Figure~\ref{figure3}. At each redshift bin, the SFRD exhibits a clear peak at a characteristic stellar mass of $\sim10^{10}$ M$_\odot$, indicating that galaxies in this mass range contribute most significantly to the cosmic star formation budget. We note that the integrated SFRD peaks at $1<z<1.5$ in our analysis, which appears to conflict with the well-established result that the cosmic SFRD peaks at $z\sim2$ \citep{Madau14}. As we will discuss later, this apparent discrepancy can be explained by observational limitations of the COSMOS2020 catalog.

The right panel of Figure~\ref{figure3} further shows the characteristic peak mass as a function of redshift. The peak mass appear to exhibit a weak positive correlation with redshift, which is tightly related to the turnover mass M$_0$ in Table~\ref{table2}. We perform a linear fit to the data and show the corresponding Pearson correlation coefficient $r$ and its $p$-value in the figure. The results reveal a weak but statistically significant linear relationship. The similar evolution of both the peak mass and turnover mass (Figure~\ref{figure2}) is consistent with the cosmic downsizing scenario, in which the most massive galaxies formed their stars and quenched earlier in cosmic history, followed by progressively less massive systems at later times \citep{Cowie96,Bundy06,Fontanot09,Mortlock11,Siudek17}. This mass-dependent evolution is likely driven by more efficient gas consumption and stronger feedback processes in massive systems, which deplete their gas reservoirs and trigger earlier quenching.

Using the same COSMOS2020 catalog, \cite{Weaver23} studied the galaxy stellar mass function at $0.2<z<7.5$, finding that the high-mass end becomes increasingly dominated by quiescent galaxies at $z<1.5$, while lower-mass galaxies remain predominantly star-forming. A further analysis of quiescent fraction revealed that it increases with increasing mass at fixed redshift and increases with cosmic time at fixed mass. The convergence of our results with those of \cite{Weaver23} provides compelling evidence for the downsizing scenario, where massive galaxies cease star formation at earlier cosmic epochs than their lower-mass counterparts.

\begin{figure}[htbp]
	\centering
    \includegraphics[width=0.5\textwidth]{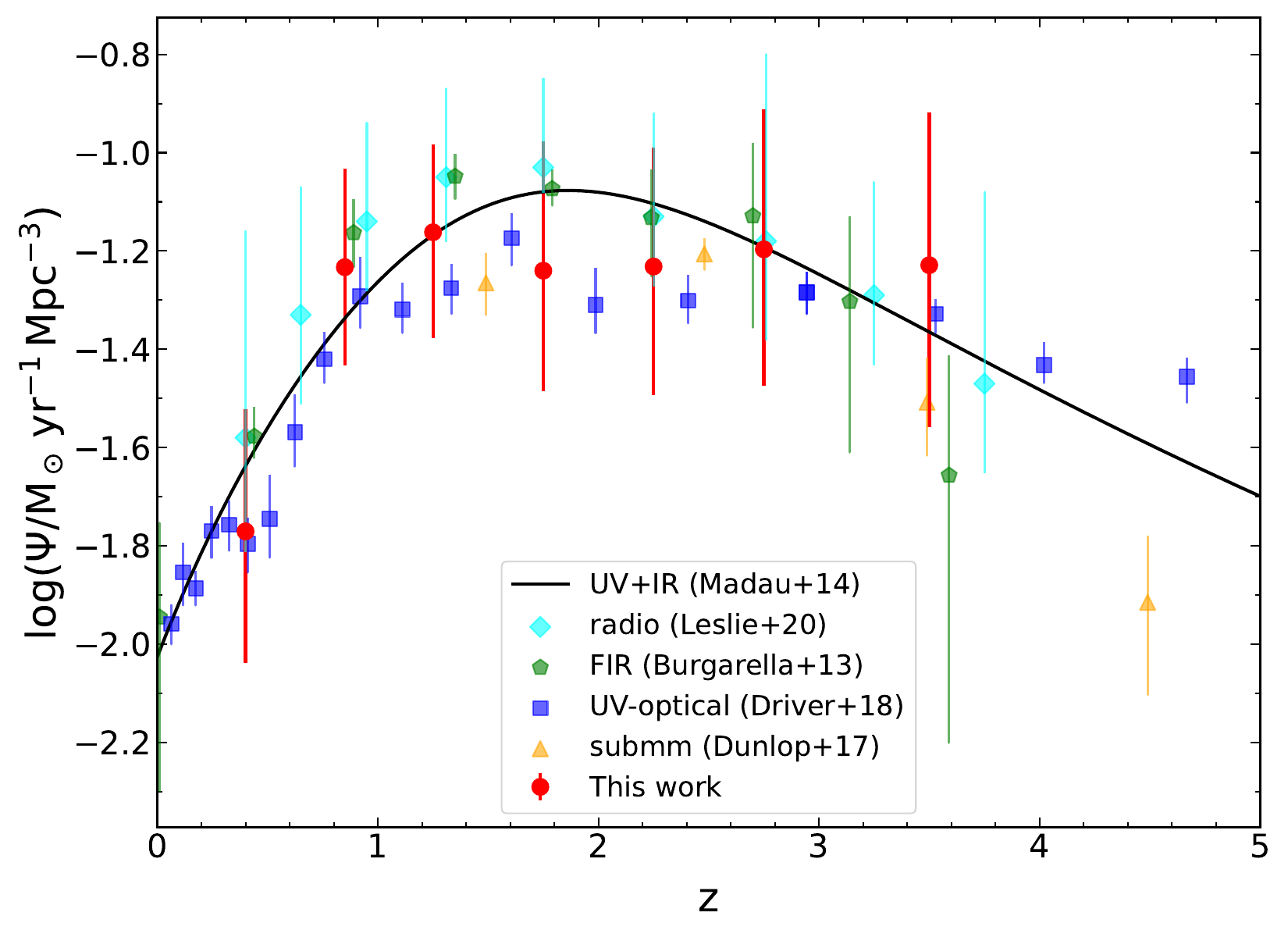}
    \caption{Redshift evolution of the cosmic star-formation rate density. Red points show our results from the COSMOS2020 catalog, compared with literature measurements using different star formation tracers. All values have been converted to a \cite{Chabrier03} initial mass function for direct comparison.}
    \label{figure4}
\end{figure}

Next, we calculate the cosmic SFRD as a function of redshift by integrating the relations in Figure~\ref{figure3} over the stellar mass range of $10^8 M_\odot - 10^{12} M_\odot$. Our result is presented in Figure~\ref{figure4}.  For comparison, we also plot SFRD measurements from the literature derived using different tracers: UV-optical \citep{Driver18}, FIR \citep{Burgarella13}, submillimeter \citep{Dunlop17}, and radio \citep{Leslie20}, along with the UV$+$IR functional form of \cite{Madau14}. 

Our results are generally consistent with the literature, except at $1.5<z<3$ where our SFRD values are systematically lower than FIR- and radio-based measurements by approximately 0.2 dex. This discrepancy reflects a fundamental limitation of UV-optical-NIR observations: they are highly susceptible to dust extinction and systematically miss or underestimate the star formation rates of heavily dust-obscured galaxies. At cosmic noon, the cosmic mean dust attenuation reaches its peak \citep{Madau14}. Consequently, a substantial population of star-forming galaxies are heavily dust-enshrouded, with UV and optical emissions severely attenuated but FIR \citep{Gruppioni20,Traina24} and radio emissions \citep{Talia21,Gentile25} remaining detectable, creating a strong observational bias favoring long-wavelength tracers. Since COSMOS2020 catalog lacks FIR and radio information, it could underestimate the contribution of dusty star-forming galaxies, leading to a ``dip'' we see at $z\sim2$. The same is true for the other UV-opical based observation by \cite{Driver18}, who used a combined optically selected dataset to study cosmic star-formation history from $0<z<5$ using SED fitting technique. Our results closely match those of \cite{Driver18} as shown in the figure. 

At lower and higher redshift, our results agree well with those in the literature, validating our methodology while highlighting the critical importance of multi-wavelength observations including FIR and radio information for complete SFRD estimates during the peak epoch of cosmic star formation.

\subsection{The role of environments on SFR-M$_\star$ relation} \label{env}
It is well established that environments have profound impacts on the star-formation activities of galaxies that vary dramatically across cosmic time. At low redshifts, dense environments like clusters tend to host a higher fraction of quiescent galaxies \citep[e.g.,][]{Dressler80, Dressler97, Goto03, Kauffmann04}. In contrast, many high-redshift studies have observed enhanced SFRs in dense environments such as protoclusters \citep[e.g.,][]{Elbaz07,Tran10,Koyama13,Shimakawa18,Shi20,
Lemaux22,Shi24}. 

However, whether the SFR-M$_\star$ relation, or the SFMS depends on environments is still under debate. While many studies found no obvious environmental dependence of the SFMS in different environments \citep{Ricciardelli14,Grossi18,Randriamampandry20,
Pharo20,Cooke23}, numerous low redshift cluster studies have reported suppressed SFRs of cluster galaxies relative to their field counterparts \citep[e.g.,][]{Couch01,Gomez03,Haines13,Paccagnella16,
Schaefer19,Old20,Finn23}.  These different findings may stem from differences in sample selection, environmental definitions, redshift ranges, and methodological approaches. In this section, we investigate in detail the environmental effects on the SFR-M$_\star$ relation using our large, homogeneous sample to provide new insights into this ongoing controversy.

To explore the environmental dependence of the SFR-M$_\star$ relation at different redshifts,  it is essential to employ a uniformly selected galaxy sample. Since the mass completeness limit varies with both redshift and galaxy type, applying different limits at each redshift would introduce artificial trends -- for instance, including lower-mass galaxies only at low-$z$ could mimic or obscure genuine environmental effects. To avoid this, we adopt the mass completeness limit of the highest redshift bin studied in this work, which is 10$^{9.26}$ M$_\odot$ for all galaxies and 10$^{9.64}$ M$_\odot$ for quiescent galaxies at $z\sim4$, and apply it uniformly across all redshift bins. This leads to a total of 147,012 galaxies in the subsample.

The environments of galaxies in this study are defined using the same methodology as detailed in \cite{Shi24}, and we refer readers to that work for a comprehensive description of the environmental measurement techniques. First, we constructed a series of narrow redshift slices and consider each slice as a 2D structure. Photometric redshift errors in COSMOS2020 catalog range from $\sim$0.01 at $z\sim0.1$ to $\sim$0.1 at $z\sim4$. To take these uncertainties into account, the width of each slice is set to be the median value of the photometric redshift uncertainty $\Delta z=z_\textrm{max}-z_\textrm{min}$ at that redshift, where $z_\textrm{max} $ ($z_\textrm{min}$) is the upper (lower) limit (68\% confidence level) of the best-fit photometric redshift. This ensures that galaxies within each slice are likely to be physically associated rather than merely projected along the line of sight. In the end, 81 slices are selected spanning from $z = 0.1-4$. We note that employing a variable slice width in redshift space means that each slice traces a different physical scale, with thicker slices at high-$z$ introducing greater line-of-sight contamination. As a consequence, a given density contrast at high-$z$ may correspond to intrinsically denser structures than the same value measured at low-$z$, which could potentially bias the interpretation of the redshift evolution of the relation. To validate our result, we also adopt an alternative approach using a fixed comoving slice width of 40 cMpc (e.g., $\Delta z \sim 0.01$ at $z = 0.1$ and $\Delta z \sim 0.05$ at $z\sim4$) and repeat the analysis. The results presented in the following sections remain unchanged, confirming that our conclusions are not biased by the choice of slicing strategy. 

We then perform Voronoi tessellation technique on each 2D slice obtained above to determine the local density value where each galaxy resides. Voronoi tessellation is a nonparametric density estimator that has been widely used in astronomy field to detect both clusters and protoclusters \citep[e.g.,][]{Ramella01, Soares11, Dey16, Shi21}. A Voronoi tessellation is a unique way of dividing a 2D distribution of galaxies into convex cells, with each cell containing only one galaxy. Then the local density of each cell is the inverse of the cell area. To determine the relative density contrast of each cell, we calculate the average density of all cells in the entire 2D plane and the density contrast of each cell is obtained by dividing the average density value. 

We classify galaxy environments based on the density contrast distribution: low-density regions correspond to the lowest 25th percentile, high-density regions to the highest 25th percentile, and the intermediate 50\% to average-density environments. We note the specific definition of environmental density threshold values do not affect our subsequent results.

\begin{figure*}
	\sidecaption
    \includegraphics[width=12cm]{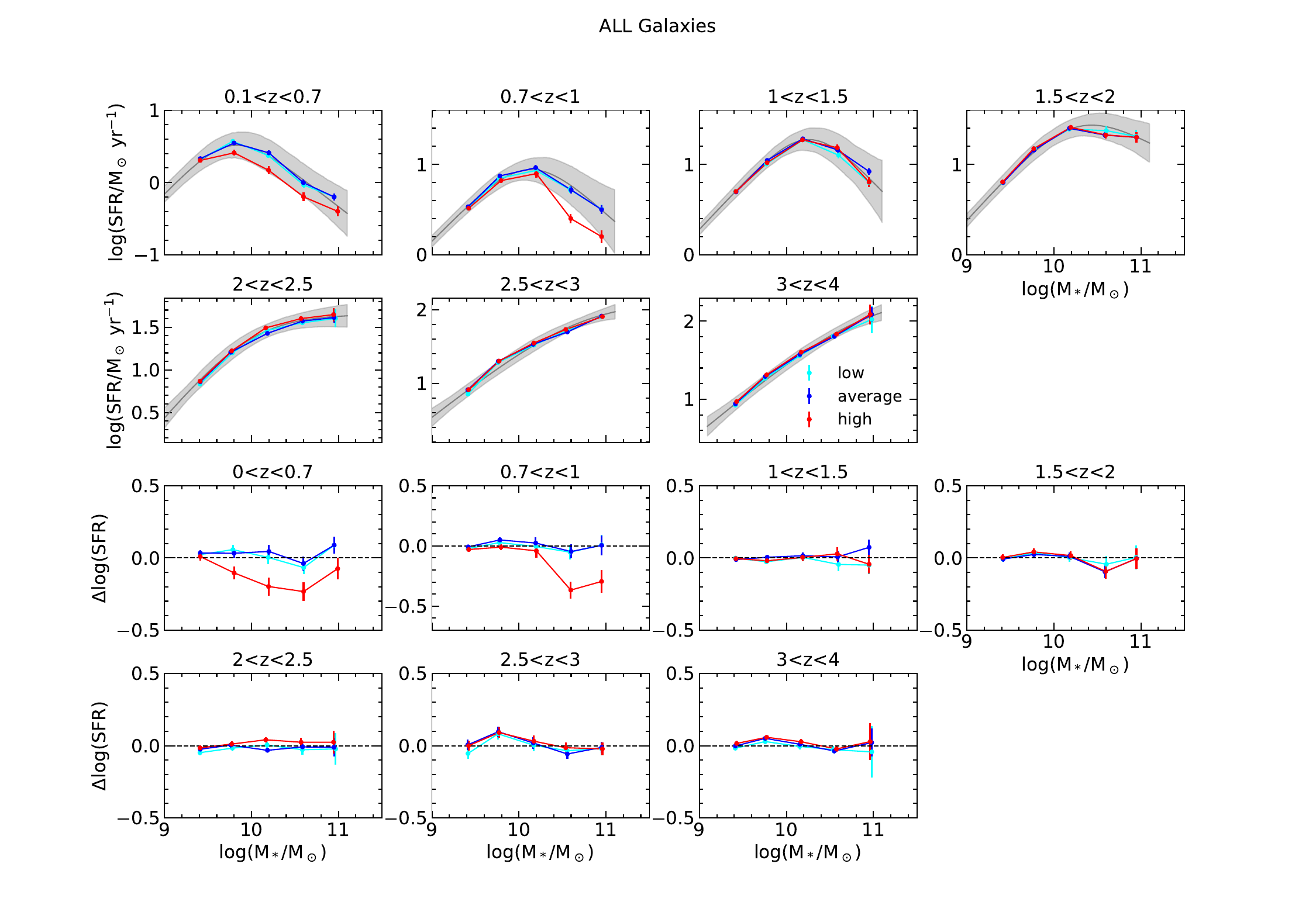}
    \caption{Environmental dependence of the SFR-M$_\star$ relation for all galaxies. The upper seven panels show the SFR-M$_\star$ relation in each redshift bin for all galaxies in different local densities as defined in the text. The best-fit relation for all galaxies in Figure~\ref{figure1} is shown as the grey line with shaded area denoting the 1$\sigma$ confidence level. The lower seven panels show the residuals calculated as the log difference between galaxies in different local environments and the best-fit relation.}
    \label{figure6}
\end{figure*}

In Figure~\ref{figure6}, we show the SFR-M$_\star$ relation for all galaxies in different local environments. It can be clearly seen that galaxies in high-density environments show suppressed SFRs at the high-mass end up to $z<1$. This suppression likely reflects a combination of mass-driven quenching of central galaxies, which dominate at high stellar masses, and the environmental quenching of satellite galaxies, whose fraction at fixed stellar mass is higher in denser environments \citep{Peng12}. At lower redshifts, the cumulative accretion of satellites over cosmic time, combined with the maturation of the intracluster medium in increasingly massive host halos, results in a larger quenched population in dense environments. In contrast, at $z>1$, no significant environmental effects are observed.

\begin{figure*}[htbp]
	\sidecaption
    \includegraphics[width=12cm]{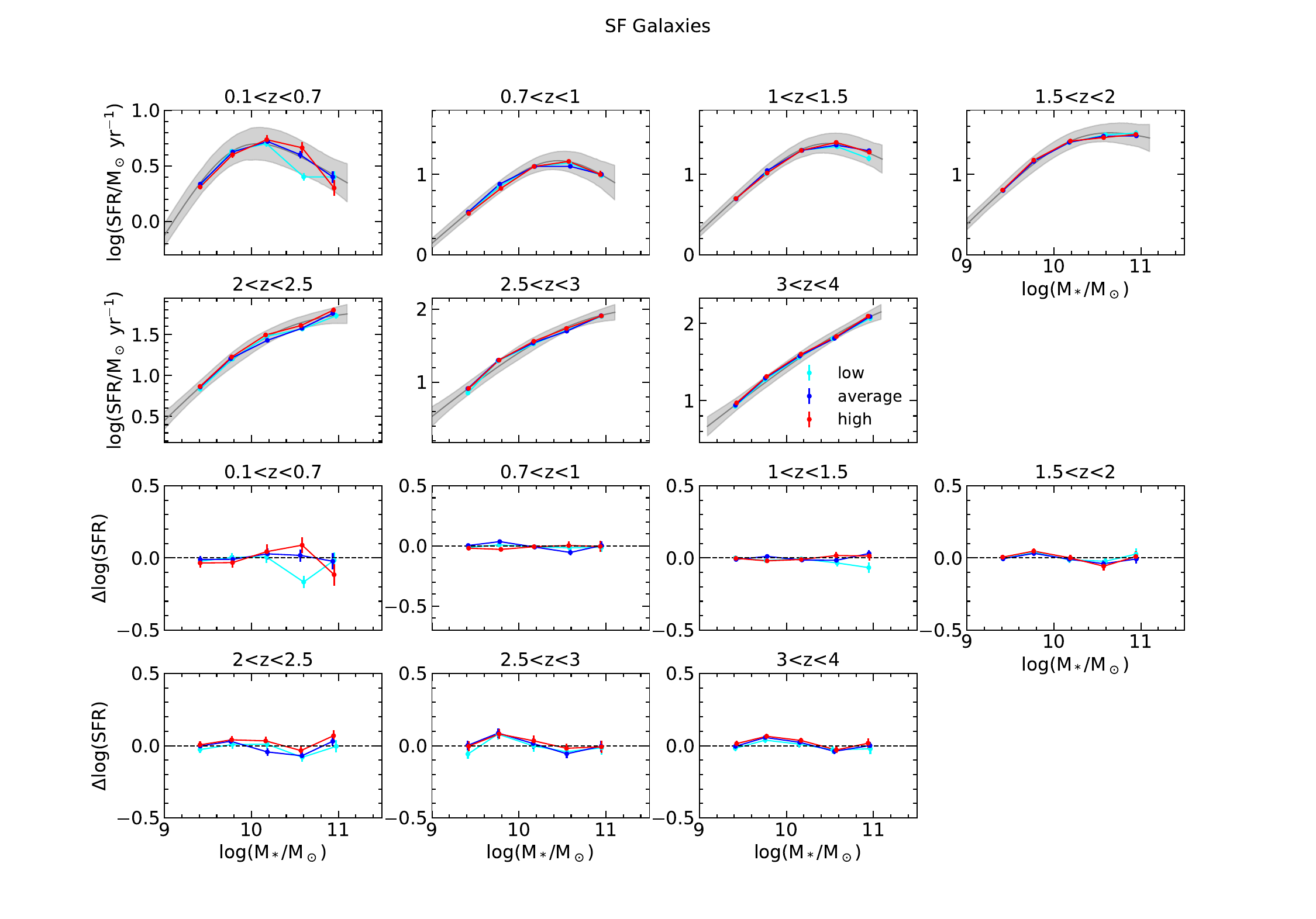}
    \caption{Same as Figure~\ref{figure6} but for star-forming galaxies. The best-fit relation for star-forming galaxies in Figure~\ref{figure1} is shown as the grey line with shaded area denoting the 1$\sigma$ confidence level.}
    \label{figure7}
\end{figure*}

It remains to be explored whether the suppression of SFRs at $z<1$ is mainly caused by the presence of quiescent galaxies or whether it represents a general trend for all galaxies (including star-forming ones). Therefore in Figure~\ref{figure7}, we show the SFR-M$_\star$ relation for star-forming galaxies only. In this case, no obvious differences are found among different environments. This suggests that for star-forming galaxies, their position on the main sequence is primarily determined by their stellar mass rather than their local environments. However, we note that  at around $\sim10^{10.5}~\textrm{M}_\odot$ at $z<0.7$, star-forming galaxies in low-density environments show tentatively lower SFRs than those in average/high-density environments, but this signal is limited to this single mass bin and not statistically significant (K-S test $p=0.06$). Larger samples are needed to confirm whether this reflects a genuine environmental dependence or simply a random statistical fluctuation.

\cite{Shi24} also used the same data and method to investigate the environmental impact on star formation activities of galaxies. They found that at $z<1$ the SFR of all galaxies decrease with increasing density of the environment, but the trend disappears when they consider only star-forming galaxies. By comparing the quenching efficiencies induced by environment and mass, they argued that environmental quenching is only relevant at $z<1$, and stellar mass may be the major driving force for quenching massive galaxies at $z<1$, dominating over environmental effects. Our results are consistent with \cite{Shi24}, suggesting that dense environments cause a more efficient assembly of massive galaxies through galaxy-galaxy interactions and mergers. Once massive galaxies form, their deep gravitational wells heat infalling gas and suppress star formation, with AGN and stellar feedback further reinforcing this quenching. In dense environments, the cumulative accretion of satellite galaxies — which are subject to additional quenching processes such as ram-pressure stripping \citep{Gunn72} and strangulation \citep{Larson80} — further builds up the quiescent population over cosmic time, making this effect most prominent at low redshifts. In contrast, star-forming galaxies roughly follow the same main sequence relation regardless of environment, showing no systematic offset in SFR at fixed stellar mass between high-density and low-density regions.  This consistency suggests that their star formation activities are mainly governed by internal processes (i.e., stellar mass-dependent gas accretion, stellar feedback, and disk stability) rather than external ones (i.e., ram-pressure stripping, galaxy interactions, or ICM heating). 

\begin{figure}[htbp]
	\centering
    \includegraphics[scale=0.5]{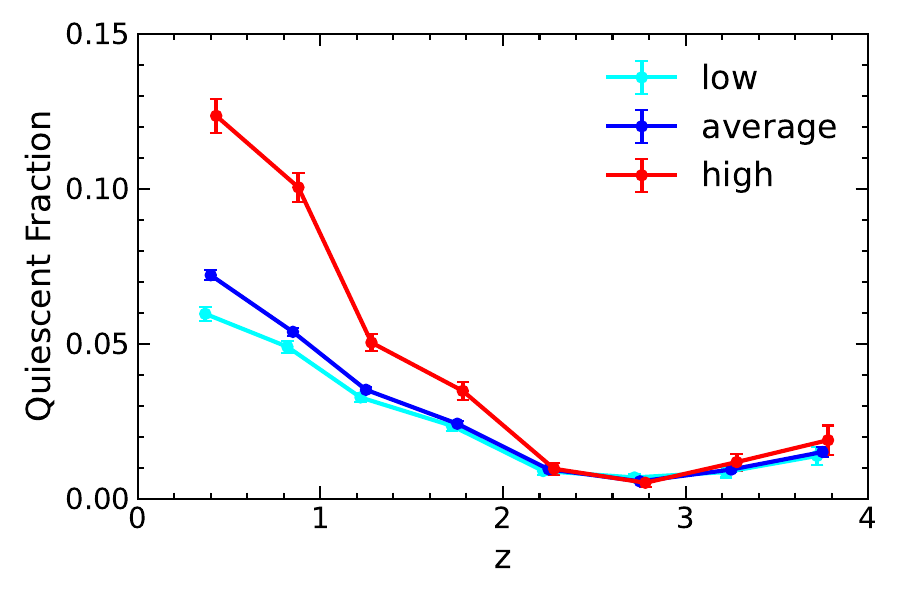}
    \caption{Quiescent fraction as a function of redshift in different environments. The error bar denotes the Poisson error.}
    \label{figureq}
\end{figure}

To investigate whether the suppressed SFR in high-density regions at $z<1$ for all galaxies is primarily due to the increasing fraction of quiescent galaxies, in Figure~\ref{figureq} we show the quiescent fraction as a function of redshift in different environments. It can be clearly seen that at $z\lesssim2$, the quiescent fraction is substantially higher in high-density environments than in average- or low-density ones, consistent with the expected signatures of environmental quenching. Therefore, dense environments such as clusters mainly increase the fraction of quiescent galaxies, leading to the declined SFRs observed at low redshifts. 

The lack of environmental effects on star-forming galaxies at all redshifts, on the other hand, suggests that environmental quenching is a rapid transformation rather than a slow starvation process. If environmental mechanisms were the dominant quenching pathway, one would expect star-forming galaxies in dense environments to exhibit systematically suppressed SFRs relative to their field counterparts, reflecting the progressive exhaustion of their gas supply. The absence of such intermediate-state galaxies in our sample implies that environmental quenching instead acts as a ``binary'' switch that completely shuts down star formation on timescales shorter than can be resolved in our redshift bins. This rapid quenching scenario is more consistent with mechanisms such as ram pressure stripping, which can remove the cold gas disk on timescales of a few hundred Myrs as galaxies traverse the dense intracluster medium \citep[e.g.,][]{Quilis00,Vollmer01,Salinas24}. The effectiveness of this process at low redshifts, coupled with its apparent inefficiency at $z \gtrsim 2$ where environmental differentiation vanishes, may reflect the evolving properties of galaxy clusters themselves: at earlier epochs, cluster potential wells were shallower and the intracluster medium less developed, rendering ram pressure stripping less effective as a quenching mechanism. This picture is corroborated by observational studies of high-redshift protoclusters, which consistently find enhanced star formation rates among member galaxies compared to field counterparts \citep[e.g.,][]{Tran10, Shimakawa18, Shi20}. By $z \lesssim 2$, however, clusters have undergone substantial virialization to develop deep potential wells and hot, dense gaseous halos capable of efficiently stripping infalling satellites of their cold gas reservoirs.

\cite{Cooke23} also investigated the role of environments on the SFMS in the COSMOS field from $0<z<3.5$, using the COSMOS2015 catalog \citep{Laigle16}. They found that star-forming galaxies show no environmental dependence at any redshift in their sample. The deeper COSMOS data employed in our study fully support their results. Our work is also consistent with \cite{Tomczak19}, who studied the environmental dependence of SFR-M$_\star$ relation at $z\sim0.9$, finding a strong decline of SFR with increasing density of the environments for all galaxies. However when star-forming galaxies are considered, the trend becomes much weaker. Similar results have been reported by \cite{Erfanianfar16}, who divided galaxies into group and field environments, and found that at low redshift group galaxies deviate towards quiescence compared to isolated galaxies, while at high redshifts this environmental effect disappears. Nevertheless, we note that \cite{Leslie20} found no significant environmental trend in the SFR-M$_\star$ relation for either star-forming or all galaxy samples using radio 3GHz data in the COSMOS field. However, since their SFR is derived from radio data,  their results suffer from larger uncertainties due to lower spatial resolution and more severe source blending, which can potentially result in non-detections of environmental effects.

\subsection{The role of morphology on SFR-M$_\star$ relation} \label{morp}
Aside from stellar mass and environment, it is well known that galaxy morphology also has significant effect on star-formation activities of galaxies \citep[e.g.,][]{Kauffmann03,Martig09, Wuyts11, Bluck14, Whitaker15, Brennan15, Tacchella19}. However, since most of the observational studies focus on the local or low-redshift Universe, it remains uncertain whether the morphology effect persists at high redshifts. To this end, we investigate the morphological trends on the SFR-M$_\star$ relation in this section. 

The galaxy morphological information used in this study are taken from two catalogs. One is the Tasca Morphology Catalog (v1.1) \citep{Tasca09}, which contains the morphological information of 237,192 sources in the COSMOS field that have Hubble Space Telescope (HST) Advanced Camera for Surveys (ACS) observations \citep{Leauthaud07} in the $\textit{F814W}$ filter. The morphological parameters are estimated using the \texttt{MORPHEUS} code \citep{Abraham07}. A training set of $\sim$500 galaxies was visually classified into early-type, spiral, and irregular morphologies, and machine-learning algorithms were then applied to classify the remaining galaxies.  We note that objects in the Tasca catalog are selected with $i$-band magnitude $<23$, substantially brighter than the COSMOS2020 limiting magnitude of $i=26$. The Tasca sample is therefore biased toward intrinsically luminous galaxies with elevated SFRs relative to the parent population. In addition to Tasca catalog, we also include the newly released COSMOS-Web galaxy catalog from COSMOS2025 data release \citep{Shuntov25}, which contains more than 700,000 sources in the COSMOS field. Galaxies are classified into three broad morphological types (early-type, spiral and irregular) using supervised machine learning technique from \cite{Huertas24}. The morphological information we employed is derived from F277W band which is the deepest in NIRCam filters. We then combine the two morphology catalogs into one, adopting the measurements from the COSMOS-Web catalog for overlapping sources. This master morphology catalog is then cross matched with our COSMOS2020 catalog using a 1$\arcsec$ radius, yielding 177,604 counterparts, among which 19,251 are early-type galaxies and 78,307 are spirals and 80,046 are irregulars. 

\begin{figure*}
	\sidecaption
    \includegraphics[width=12cm]{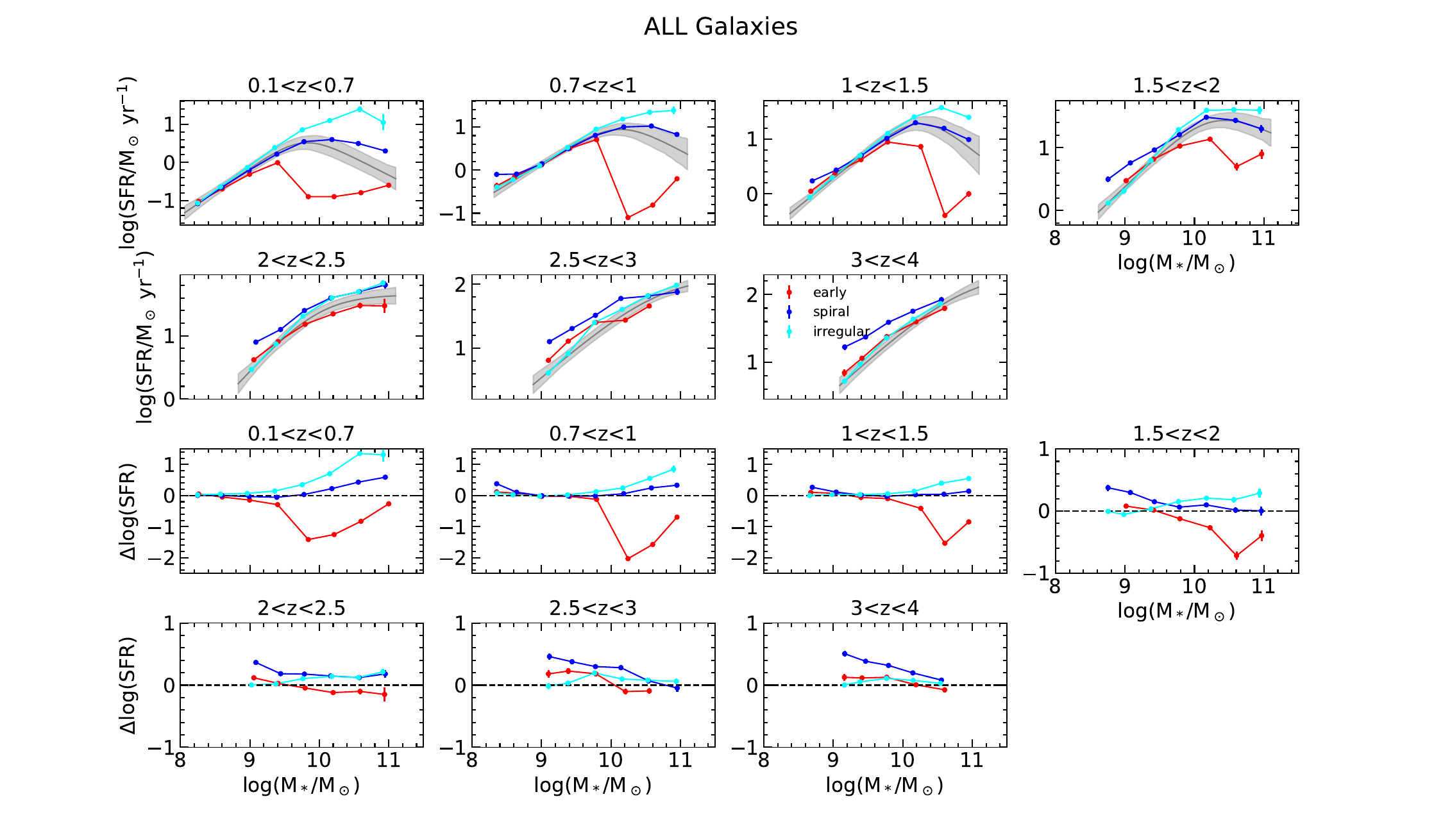}
    \caption{SFR-M$_\star$ relation for all galaxies with different morphology types. The upper seven panels show the SFR-M$_\star$ relation in each redshift bin for all galaxies of different morphologies as defined in the text. The best-fit relation for all galaxies in Figure~\ref{figure1} is shown as the grey line with shaded area denoting the 1$\sigma$ confidence level. The lower seven panels show the residuals calculated as the log difference between galaxies of different morphologies and the best-fit relation.}
    \label{figure8}
\end{figure*}

Figure \ref{figure8} shows the morphological trends for all galaxies at different redshifts. The best-fit relation of all galaxies from Figure~\ref{figure1} is shown for comparison. The fact these galaxies are skewed above the best-fit relation is due to the selection bias mentioned earlier.  It is clear that up to $z<2$, early-type galaxies are strongly quenched compared to late-type galaxies, exhibiting systematically lower SFRs at fixed stellar mass, with this suppression being particularly pronounced at high masses ($\log M_\star / M_\odot \gtrsim 10$). In contrast, late-type galaxies generally have enhanced star formation, with irregulars exhibit the highest SFRs among the three types. This clear segregation  demonstrates that galaxy morphology plays a fundamental role in regulating star formation. It is also noted at $z>2$, spiral galaxies appear to have elevated SFRs compared to early types and irregulars. This is also most likely attributable to a selection effect. Since the machine learning technique employed in COSMOS-WEB catalog requires that galaxies classified as disk-dominated have visually identifiable disk features. At high redshift, this would preferentially select only the most prominent, high surface brightness disks, biasing the disk-dominated sample toward more actively star-forming systems while fainter disks are likely classified as irregular.

\begin{figure*}
	\sidecaption
    \includegraphics[width=12cm]{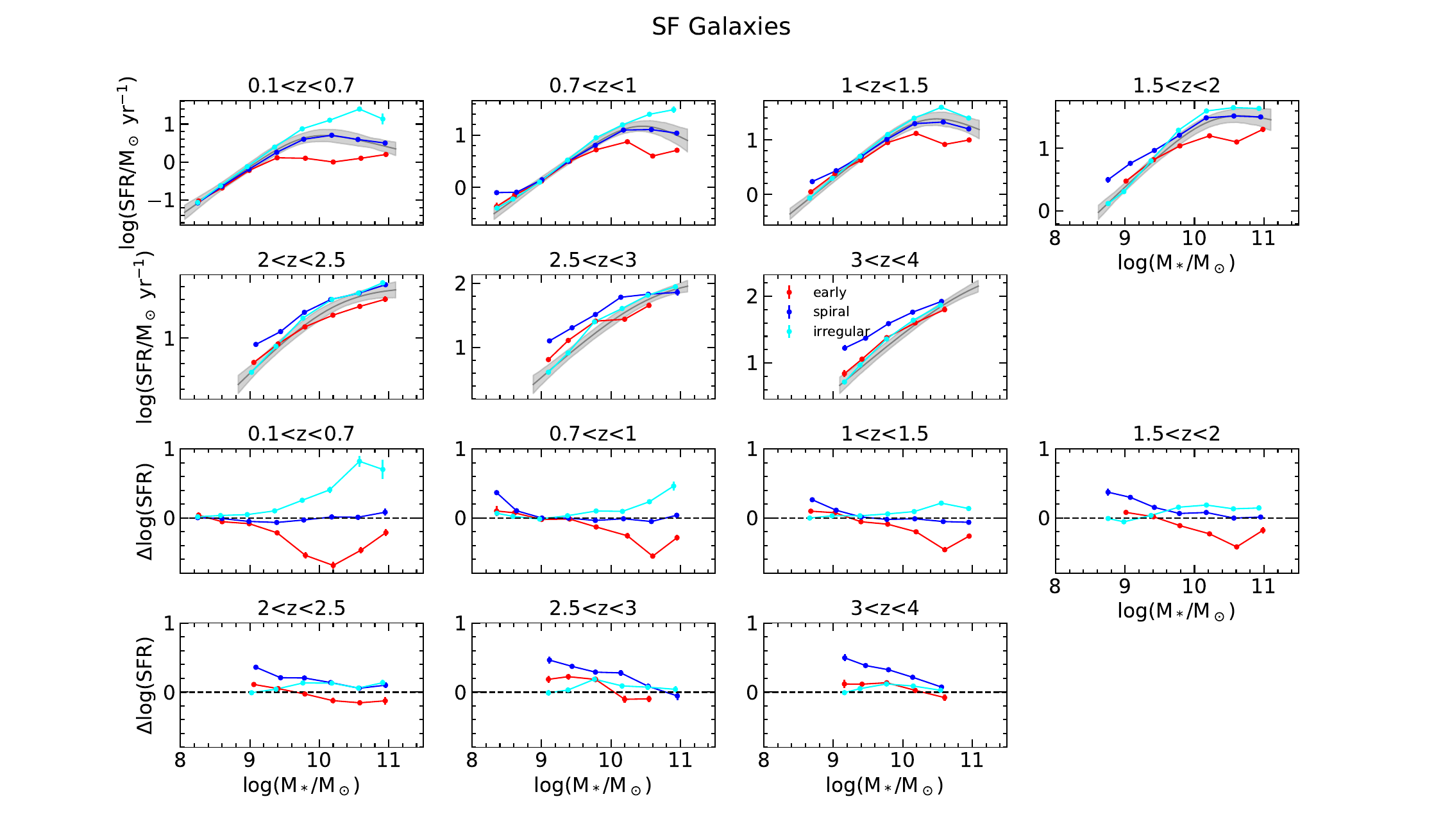}
    \caption{Same as Figure~\ref{figure8} but for star-forming galaxies. The best-fit relation for star-forming galaxies in Figure~\ref{figure1} is shown as the grey line with shaded area denoting the 1$\sigma$ confidence level.}
    \label{figure9}
\end{figure*}

Furthermore, Figure \ref{figure9} presents the analysis for the star-forming galaxy subsample. Similar to all galaxies, star-forming galaxies also show a systematic correlation with morphological type on the SFMS. Specifically, early-type galaxies exhibit the lowest SFRs up to $z<2$, followed by spirals and irregulars. This suggests that morphological quenching operates effectively even within the star-forming galaxy population at $z<2$, indicating that galaxy structure significantly regulates star formation efficiency regardless of whether galaxies are classified as active or passive. 

The strong quenching of early-type galaxies in both the full sample and the star-forming subsample is most likely driven by the presence of a dominant bulge component. In this scenario, the centrally concentrated bulge would suppress star formation by stabilizing the gas disk against fragmenting, even when significant cold gas remains \citep{Martig09}. This stabilization increases the disk's Toomre stability parameter, preventing the collapse of gas into star-forming clumps. Recent observations from low redshift ($0<z<0.5$) \citep{Lesniewska23} and intermediate redshift ($0.5<z<1.5$) \citep{Lu21} both support this quenching scenario. Alternatively, since the growth of supermassive black holes is tightly correlated with bulge growth through galaxy mergers, AGN feedback can also drive quenching either through direct heating and expulsion of gas from the central bulge region, or indirect starvation by preventing cold gas accretion from the halo through shock heating and turbulent energy injection \citep{Croton06,Fabian12}.

\begin{figure}[htbp]
	\centering
    \includegraphics[scale=0.5]{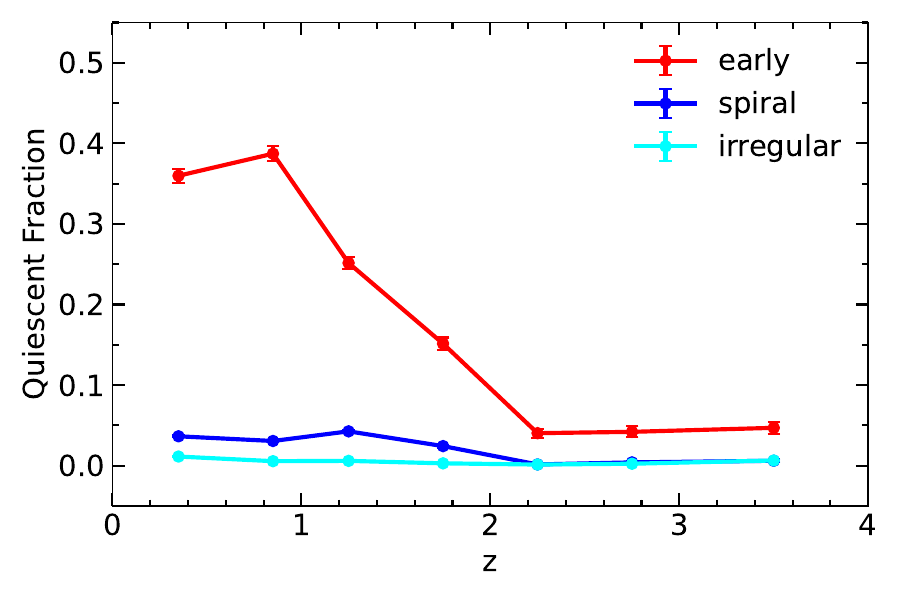}
    \caption{Quiescent fraction as a function of redshift for different morphological types. The error bar denotes Poisson error.}
    \label{figureqm}
\end{figure}

Of particular interest is the finding that while environment has minimal influence on the SFRs of star-forming galaxies (Figure~\ref{figure7}), morphology demonstrates a significant and systematic effect on star formation efficiency of this same population (Figure~\ref{figure9}). Indeed, morphology appears to exert a stronger influence on quiescence than environment. In Figure~\ref{figureqm} we show the quiescent fraction as a function of redshift and morphology types. Early-type galaxies display substantially elevated quiescent fractions compared to their late-type counterparts at $z < 2$, reaching a maximum of approximately 50\% at $z\sim0.8$ before declining steeply toward higher redshifts. In contrast, spiral and irregular galaxies maintain uniformly low quiescent fractions ($<$5\%) throughout the redshift range examined. This implies that the structural transformation accompanying bulge formation is more fundamentally linked to quenching than external environmental processes. 

The distinct quenching behaviors associated with environment and morphology may reflect their different quenching timescales. Environmental quenching is generally described by a `delayed-then-rapid' scenario \citep{Wetzel13}, in which satellite galaxies evolve largely unaffected for a few Gyr after infall before undergoing rapid quenching on timescales of $\lesssim$1 Gyr, with shorter total timescales at higher redshifts and stellar masses \citep{Foltz18,Walters22}. In this context, galaxies are either still in the delay phase and actively star-forming, or have already undergone rapid quenching and become quiescent, leaving few systems in intermediate states. In contrast, morphology exerts a strong influence on star formation as can be seen in both the full and star-forming subsample up to $z<2$, indicating that morphological transformation contributes to quenching in a more gradual and sustained manner compared to rapid environmental processes. This finding depicts an evolutionary picture where morphological transformation and star formation quenching are coupled but potentially separable processes. Galaxies can undergo structural changes that reduce their star formation efficiency while remaining active, as evidenced by the morphological dependence within star-forming systems. However, complete quenching requires additional mechanisms, either internal (i.e., mass) or external (i.e., environment), that remove or heat the remaining gas beyond what structural changes alone accomplish. This is supported by the observational study of \cite{Fang13}, who analyzed a sample of central galaxies in SDSS and found that bulge growth is a necessary but insufficient condition for quenching. 
Their results indicate that halo-related processes -- such as virial shock heating that prevents cold gas accretion -- must also operate to ensure complete and sustained quenching. Therefore, transformation from actively star-forming disk galaxies to passive early-type systems may involve multiple stages, with morphological transformation preceding the final departure from the SFMS \citep{Barro13,Lang14,Bremer18}. 

Our findings are fully consistent with \cite{Leslie20}, who conducted a comprehensive analysis of morphological trends of galaxies on the SFMS in the COSMOS field and found that bulge-dominated galaxies exhibit systematically lower star formation rates than disk-dominated systems at $z<1$. Similar results have been reported by \cite{Dimauro22}, who studied a sample of massive galaxies at $0<z<2$ on the SFR-M$_\star$ plane, finding galaxies with minimal bulge components ($B/T<0.2$) are all on the SFMS while galaxies with significant bulges ($B/T>0.2$) are responsible for the bending of main sequence at high-mass end.  In the local Universe, \cite{Cano19} investigated in detail both the global and spatially resolved SFR-M$_\star$ relation using the data from Mapping Nearby Galaxies at APO (MaNGA) MPL-5 survey, finding that earlier type galaxies tend to form stars at lower rates at the $\sim$kpc scales. All in all, this spatial segregation within the main sequence locus reflects the systematic differences in star formation efficiency between different morphological types, with bulge-dominated early-type galaxies forming stars less efficiently at fixed stellar mass due to their less favorable structural conditions for sustained star formation.

\subsection{Comparison with literature} \label{compare}
\begin{figure*}[htbp]
	\centering
    \includegraphics[width=1\textwidth]{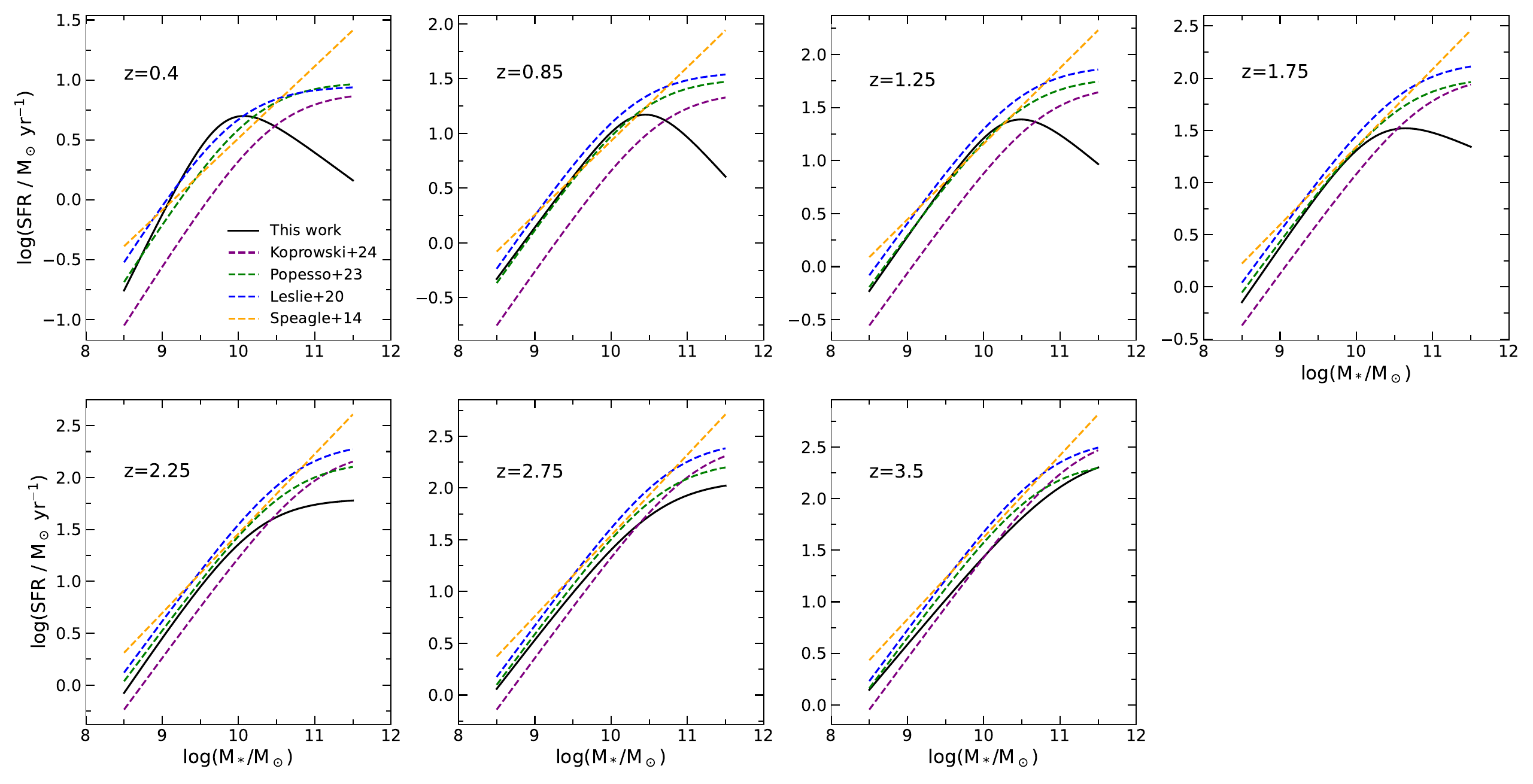}
    \caption{Star-forming main sequence relation compared with literature at different redshifts. A notable high-mass turnover can be seen in this work, especially at low redshifts.}
    \label{figure5}
\end{figure*}

In this section we compare our SFR-M$_\star$ relation of star-forming galaxies (i.e., SFMS) with those in the literature at similar redshifts, which is shown in Figure~\ref{figure5}. It is clear that at low redshifts ($z<2$), our results reveal a significant high-mass turnover around $\sim10^{10.5}$~M$_\odot$, compared with other studies. Beyond this turnover mass, our main sequence shows a steep decline compared to literature relations, with the offset reaching $\sim$0.5 dex at $10^{11}$ M$_\odot$. At high redshift ($z>2$), our relation becomes more consistent with the literature, showing an asymptotic behavior toward a maximum SFR at the high-mass end. However, a slightly steeper decline can still be seen at high-mass end.

Among these literature relations, \cite{Speagle14} simply used a linear fit to describe the main sequence while the other three adopted the \cite{Lee15} functional form. As discussed in Section~\ref{param}, neither a simple linear relation nor the asymptotic \cite{Lee15} form adequately captures the behavior of our COSMOS2020 data, which exhibits a pronounced decline at high masses, particularly at $z<2$. We therefore employed a modified functional form (Equation~\ref{eq1}) that allows for both turnover and subsequent decline at the high-mass end. 

We remind the readers that we are only using star-forming galaxies in this case, so the steep decline of main sequence at high-mass end cannot be explained by higher fraction of massive quiescent galaxies. We speculate that if this represents a genuine physical signature rather than an observational artifact, it demonstrates COSMOS2020's capability to detect massive star-forming galaxies undergoing suppressed star formation. These ``Green Valley'' galaxies \citep{Salim14} are relatively rare due to their rapid transition timescales ($\sim$1 Gyr, \cite{Bremer18}) from star-forming to quiescent. Together with their faint magnitudes, these galaxies are generally missed in previous surveys with insufficient depth and/or limited survey volumes.

Alternatively, the discrepancy could stem from different definition of star-forming galaxies. As described in Section~\ref{sec2}, COSMOS2020 employed  $NUVrJ$ color selection to separate star-forming and quiescent galaxies. While \cite{Leslie20} uses the same criterion, \cite{Speagle14} and \cite{Popesso23} used mixed methods for their compiled literature data, and \cite{Koprowski24} adopted $UVJ$ color criteria instead. These differences in star-forming galaxy selection may contribute to systematic offsets in the derived main sequence relations, particularly if different criteria capture different populations of transitional or green valley galaxies. However, since the \cite{Leslie20} relation still differs significantly from ours despite using the same $NUVrJ$ criterion, and since they demonstrated that their results remain nearly the same whether adopting $NUVrJ$ or $UVJ$ selection, the pronounced decline at the high-mass end in our relation is unlikely  resulted solely from differences in galaxy classification criteria. Nevertheless, as a robustness test, we also adopted a different selection criterion of sSFR$>10^{-11}$~yr$^{-1}$ to define star-forming galaxies and present our main sequence results in the appendix. As can be seen, these two different selection methods yield consistent results, confirming our results are not sensitive to specific galaxy classification method.

We note that COSMOS2020 catalog does not include FIR/radio observations, whereas the other compared studies incorporate FIR or radio data. As discussed in Section~\ref{SFRD}, the lack of FIR/radio information in COSMOS2020 could potentially underestimate the dust obscured SFR and lead to the decline in our main sequence relation. However, if dust bias were the primary cause, the decline should be most pronounced at cosmic noon ($z\sim2$) where dust obscuration peaks, rather than at $z<1$ where we actually observe the strongest effect. This argues against dust bias as the dominant explanation for our high-mass turnover. Therefore, we conclude that the primary driver of the high-mass turnover is genuine physical suppression of star formation in massive galaxies, which becomes increasingly important at lower redshifts consistent with the downsizing scenario.

Finally, we summarize the primary causes for the high-mass turnover of SFMS that is particularly significant at low redshifts. In Sections~\ref{env} and~\ref{morp}, we demonstrated that while environment has minimal impact on the main sequence relation across cosmic time, morphology exerts significant influence on the high-mass turnover. As discussed in Section~\ref{morp}, this ``morphological quenching'' is due to the presence of substantial bulge components in early-type galaxies. The development of bulge component is more favorable in massive systems driven by major mergers and violent disk instabilities, leading to the suppression of star formation at high-mass end. As galaxies become increasingly disk-dominated at higher redshifts, the high-mass turnover progressively weakens. Therefore, ``morphological quenching'' can naturally explain both the mass-dependence and redshift evolution of the main sequence turnover. This further suggests that galaxy structure serves as the primary regulator of star formation efficiency in active galaxies, while environment may act catastrophically to remove gas supplies and drive complete quenching.

\section{Summary and conclusion}\label{sum}
In this paper, we study the evolution of SFR-M$_\star$ relation out to $z=4$ using a mass-complete sample of $\sim$290k galaxies from the latest COSMOS2020 catalog, investigating the roles of environment and morphology. Our main findings are summarized below:

1. Similar to previous studies, we found the SFR-M$_\star$ relation exhibits a prominent turnover at high-mass end that cannot be described by a simple power-law. However, our analysis reveals a key difference: at $z<2$, the relation shows a significant decline beyond the turnover mass, rather than the asymptotic flattening reported in other works. To capture this behavior, we employ a modified functional form based on \cite{Lee15} that includes a negative high-mass slope parameter, providing a physically motivated and improved fit to the COSMOS2020 data.

2. We examine the redshift evolution of the best-fit parameters of our fitting form of SFR-M$_\star$ relation. The characteristic turnover mass and peak SFR increase monotonically with redshift, consistent with the cosmic downsizing scenario, whereas the low and high-mass slopes show no significant redshift dependence.

3. We measure the cosmic SFR density by integrating our best-fit SFR-M$_\star$ relation over the galaxy stellar mass function obtained by \cite{Weaver23}. The resulting SFRD peaks at $1<z<1.5$, lower than the canonical peak at $z\sim2$ (cosmic noon) reported in studies incorporating far-infrared and/or radio data. This can be attributed to COSMOS2020's reliance on optical-NIR photometry, which systematically underestimates SFRs in heavily dust-obscured galaxies that are particularly abundant at cosmic noon.

4. Environmental effects manifest differently for all galaxies (star-forming and quiescent) versus star-forming galaxies alone. For all galaxies, we find declined SFRs at $z<1$ in high-density regions especially at high-mass end, while at $z>1$ no apparent signature is observed. In contrast, this environmental dependence vanishes when examining only star-forming galaxies at any redshift. This suggests that environmental quenching acts as a rapid transformation that increases the quiescent fraction at low redshifts, rather than a slow starvation mechanism that gradually reduces star formation efficiency in active galaxies.

5. Galaxy morphology strongly influences star formation efficiency for all galaxies and star-forming galaxies in a similar way: at $z<2$, early-type galaxies have the lowest SFRs especially at high-mass end, followed by spirals and irregulars; while at $z>2$, this morphological segregation disappears. The presence of this morphological dependence even in star-forming galaxies suggests that structural properties such as the bulge components modulate star-formation efficiency in a continuously manner, gradually suppressing star formation before  complete quenching occurs.

6. We compare our SFR-M$_\star$ relation of star-forming galaxies with other studies at comparable redshifts. Among these studies, our relation exhibits a pronounced decline at high masses, particularly at $z<2$.  This enhanced suppression cannot be explained by increased fraction of quiescent galaxies, selection effect or dust bias. We attribute it to COSMOS2020's ability to detect faint and massive star-forming galaxies that are undergoing suppressed star-formation, that previous surveys generally missed due to insufficient depth and/or limited survey volumes. The high-mass decline is most likely due to the growth of the bulge components, even though galaxies are still active overall.

\begin{acknowledgements}
We thank the anonymous referee for a careful review and helpful comments that improved this paper. K.S. also thanks Nobunari Kashikawa for valuable discussions. K.S. acknowledge the funding from Fundamental Research Funds for the Central Universities under Grant No. SWU-KR22035. 
\end{acknowledgements}

%
\bibliographystyle{aa}
\bibliography{swu}

\begin{appendix}




\onecolumn
\section{Defining star-forming galaxies}
In principle, the shape of the SFMS is affected by how the star-forming galaxies are defined. While many studies employ color-based criteria such as $UVJ$ diagram \citep{Muzzin13} or $NUVrJ$ diagram \citep{Ilbert13}, others adopt thresholds based on specific SFR \citep{Fontanot09} or main sequence offset \citep{Donnari19}. It has been suggested that the high-mass turnover could be partially explained by different selection methods in usage \citep{Pearson23}. To test the robustness of our results,  we compare the main sequence derived using original $NUVrJ$ selection criterion with an alternative based on sSFR $>10^{-11}$ yr$^{-1}$.

\begin{figure*}[htbp]
	\centering
    \includegraphics[width=0.5\textwidth]{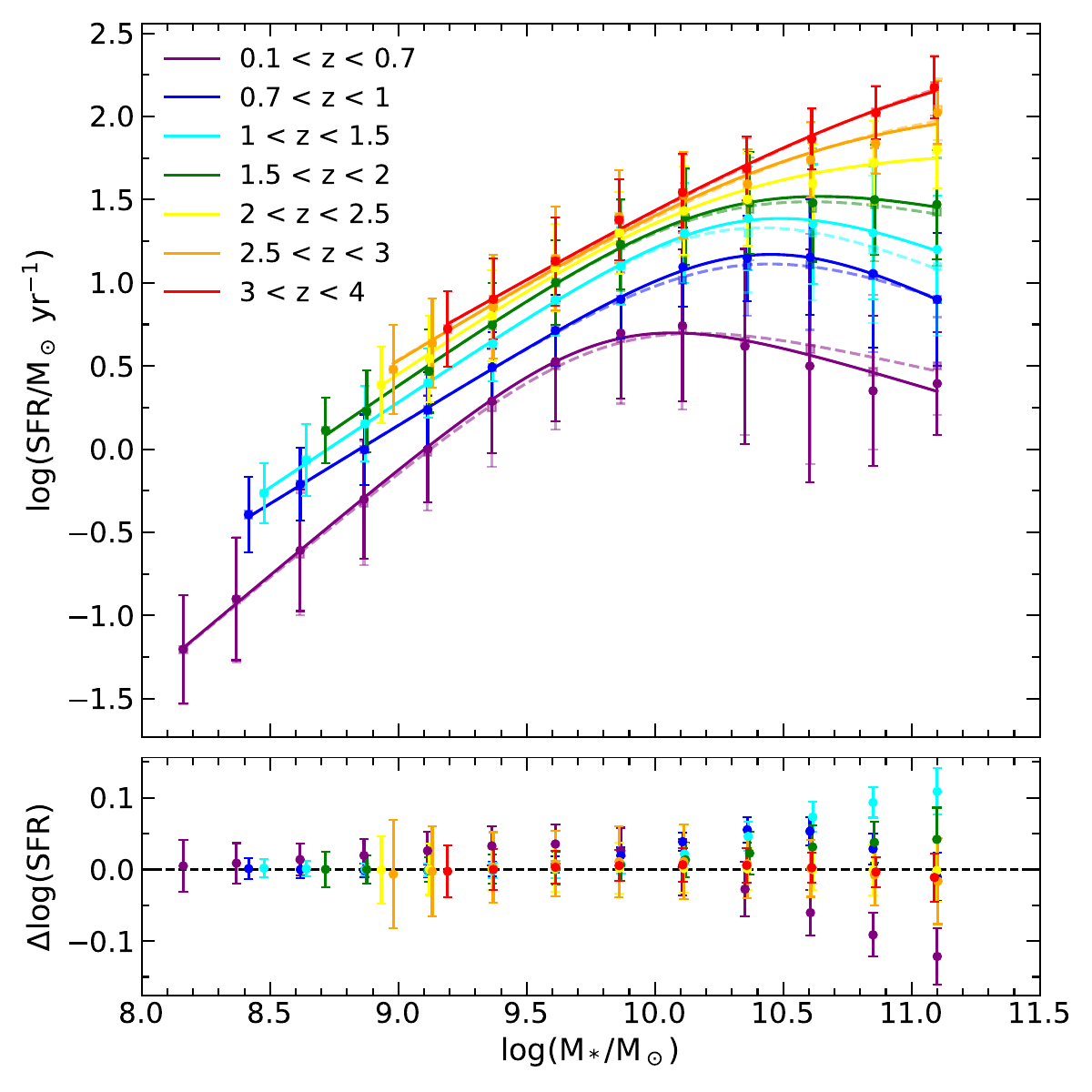}
    \caption{Star-forming main sequence relations derived using two different selection criteria for star-forming galaxies. The solid lines are the original $NUVrJ$ method used in the main text, while dashed lines represent the alternative sSFR $>10^{-11}$ yr$^{-1}$ criterion. The bottom panels show the residuals, calculated as the difference between the two relations. Both methods yield consistent results with systematic differences below 0.1 dex.}
    \label{figure10}
\end{figure*}

Figure~\ref{figure10} shows that the two selection criteria yield consistent main sequence relations, with largest offset being $\lesssim0.1$ dex across the full mass range, which is smaller than the typical main sequence scatter ($\sim$0.3 dex; \cite{Speagle14}).  The largest offsets appear at $z<1.5$ for massive galaxies ($>10^{10}$~M$_\odot$), while at $z>1.5$ the two selection methods produce nearly identical relations. Therefore, we conclude that the high-mass decline is a genuine feature in massive star-forming galaxies undergoing or approaching quenching, regardless of the specific selection method.

\end{appendix}
\end{document}